**An Eye for a Treat: Human Gazing Modulates Begging by Free-ranging Dogs**

Sourabh Biswas[1], Srijaya Nandi[1], Tuhin Subhra Pal[1], Aesha Lahiri[1], Anamitra Roy[1], Hindolii Gope[1], Kalyan Ghosh[1] and Anindita Bhadra[1]*

**Affiliations**

[1]Behaviour and Ecology Lab, Department of Biological Sciences, Indian Institute of Science Education and Research Kolkata, Mohanpur, West Bengal, India

sb18rs107@iiserkol.ac.in, sn21rs025@iiserkol.ac.in, tsp22rs044@iiserkol.ac.in, aeshalahiri@gmail.com, ar23rs018@iiserkol.ac.in, hindolii@yahoo.com, kg23rs078@iiserkol.ac.in, abhadra@iiserkol.ac.in

**ORCID ID**

Sourabh Biswas: https://orcid.org/0000-0001-6187-8106

Srijaya Nandi: https://orcid.org/0000-0002-9462-2374

Tuhin Subhra Pal: https://orcid.org/0009-0002-8953-9485

Aesha Lahiri : https://orcid.org/0000-0003-0605-3827

Anamitra Roy: https://orcid.org/0000-0002-3406-1667

Hindolii Gope: https://orcid.org/0009-0006-1355-8347

Kalyan Ghosh: https://orcid.org/0000-0002-2602-8912

Anindita Bhadra: https://orcid.org/0000-0002-3717-9732

*Corresponding Author

Prof. Anindita Bhadra

Behaviour and Ecology Lab

Department of Biological Sciences

Indian Institute of Science Education and Research Kolkata

Mohanpur Campus, Mohanpur, Nadia – 741246, West Bengal, India

Phone: +91 33 6136 0000 ext. 1223

Email: abhadra@iiserkol.ac.in

**Abstract**

Interspecific communication plays a critical role in mediating human–animal interactions, particularly in contexts involving access to anthropogenic resources. This study investigates the influence of human gazing on the begging strategies of free-ranging dogs in urban and peri-urban environments. Begging behaviour, commonly observed in dogs seeking food from humans, offers insights into their behavioural flexibility and cognitive attunement to human social cues. We observed 650 adult dogs in both solitary and group settings to assess how social context shapes the expression of begging behaviour in free-ranging dogs. Our findings indicate that solitary dogs beg more frequently than those in groups, and that female dogs exhibit higher rates of begging, predominantly through passive strategies. Moreover, dogs modulate their active begging in response to subtle variations in human gazing and food availability. These results suggest that passive begging is influenced by stable situational factors such as sex and group composition, while active begging is more responsive to immediate environmental cues, including human attentional state. Collectively, our findings highlight the social competence and behavioural plasticity of free-ranging dogs in navigating interspecies interactions, and contribute to a broader understanding of how communicative strategies evolve in response to social and ecological pressures.

## 1. Introduction

Interspecific communication offers a valuable framework for examining the evolutionary underpinnings of social cognition, particularly in contexts shaped by prolonged interspecies interactions. The ability to attend to and interpret the attentional states of others—most notably through cues such as gaze direction—has been identified as a key component of complex social behaviour across a wide range of taxa (Emery, 2000). Gaze sensitivity not only facilitates coordination and cooperation but also underlies competitive dynamics in both conspecific and heterospecific interactions. While this capacity has been documented in several lineages including primates (Kaminski et al., 2004), corvids (Bugnyar et al., 2004), and canids (Werhahn et al., 2016), its functional relevance is particularly pronounced in species that coexist closely with humans. In such systems, anthropogenic environments exert strong selective pressures that may favour individuals capable of effectively leveraging human attentional cues to secure key resources, particularly in the absence of formal human provisioning.

Among species that live in close association with humans, domestic dogs (*Canis lupus familiaris*) provide an exceptional model for investigating how animals interpret and respond to human social cues. The domestication of dogs has led to significant changes in their social and cognitive traits, resulting in a heightened sensitivity to human communicative signals compared to even our closest primate relatives (Miklósi et al., 2005; Hare & Tomasello, 2005). Dogs readily use human-given cues such as body orientation, head direction, and eye visibility to guide their behaviour in social and problem-solving contexts (Udell et al., 2010). While pet dogs in controlled environments have been the focus of much of this research, free-ranging dogs, which constitute the majority of the global dog population (Boitani & Ciucci, 1995), navigate more dynamic and unpredictable human interactions. They provide a unique and ecologically valid opportunity to study the evolution and deployment of interspecies communication strategies in real-world settings.

Begging behaviour, a common food-acquisition strategy employed by domestic and free-ranging dogs alike, offers a particularly informative context in which to examine these processes. This behaviour is sensitive to social and environmental contingencies, including human attentional state, availability of resources, and prior reinforcement history (Miklósi, 2014). Free-ranging dogs, who rely heavily on human-generated food sources but lack direct ownership or provisioning, must constantly negotiate access to resources through nuanced social strategies. In this context, sustained gazing—or the use of direct eye contact—may function as a salient communicative signal capable of eliciting human attention and increasing the likelihood of receiving food. Though dogs' sensitivity to broad attentional cues is well documented, fewer studies have explored the use and effectiveness of gazing specifically in free-ranging populations, where competition, unpredictability, and environmental noise are high.

Recent findings suggest that dogs attend preferentially to the eye region of human faces (Somppi et al., 2017), and that direct gaze from humans can elicit affiliative and attention-seeking responses (Nagasawa et al., 2015). However, the interpretation and deployment of gaze signals are context-dependent. In threatening or unfamiliar situations, dogs may avoid eye contact, whereas in cooperative or affiliative settings—such as food solicitation—gaze may play a crucial role in reinforcing social bonds and facilitating communication. This behavioural flexibility likely reflects both evolutionary history and ontogenetic experience, and may vary according to individual traits such as sex and social context.

Indeed, social context is likely to be a major determinant of the strategic use of gaze during begging. In solitary settings, dogs may be able to direct their communicative behaviour more precisely toward a human, thereby maximizing the efficacy of cues such as gaze. Conversely, in group settings, competition for human attention may necessitate more conspicuous or persistent signalling (Zajonc, 1965). The presence of conspecifics could either inhibit or amplify gazing behaviour depending on dominance hierarchies, affiliative bonds, or behavioural plasticity. Despite its theoretical importance, the interaction between gazing and social context in free-ranging dog populations remains underexplored.

This study investigates how human gazing influences begging success in free-ranging dogs, with a focus on how this behaviour is modulated by social context—specifically, whether dogs are alone or in a group. We examine whether the duration of a dog's gaze predicts its success in eliciting food and whether these patterns differ depending on the presence of conspecifics. We hypothesize that (i) gazing is more effective when dogs are solitary, as individuals can more directly target human attention without interference, and (ii) in group contexts, increased competition may drive dogs to intensify

## 2. Methodology

### 2.1 Study Sites and Subjects

This study was conducted between January and November 2023 across 14 urban and peri-urban locations in the Nadia and North 24 Parganas districts of West Bengal, India. Study sites included Kalyani, Jaguli, Barrackpore, Naihati, Balindi, Chakdah, Shyamnagar, Anandanagar, Birohi, Gayeshpur, Kanchrapara, Kataganj, Narayanpur, and Saguna (Figure 1). These locations were selected to represent a diverse range of free-ranging dog populations exposed to varying degrees of human interaction, yet comparable in human population density.

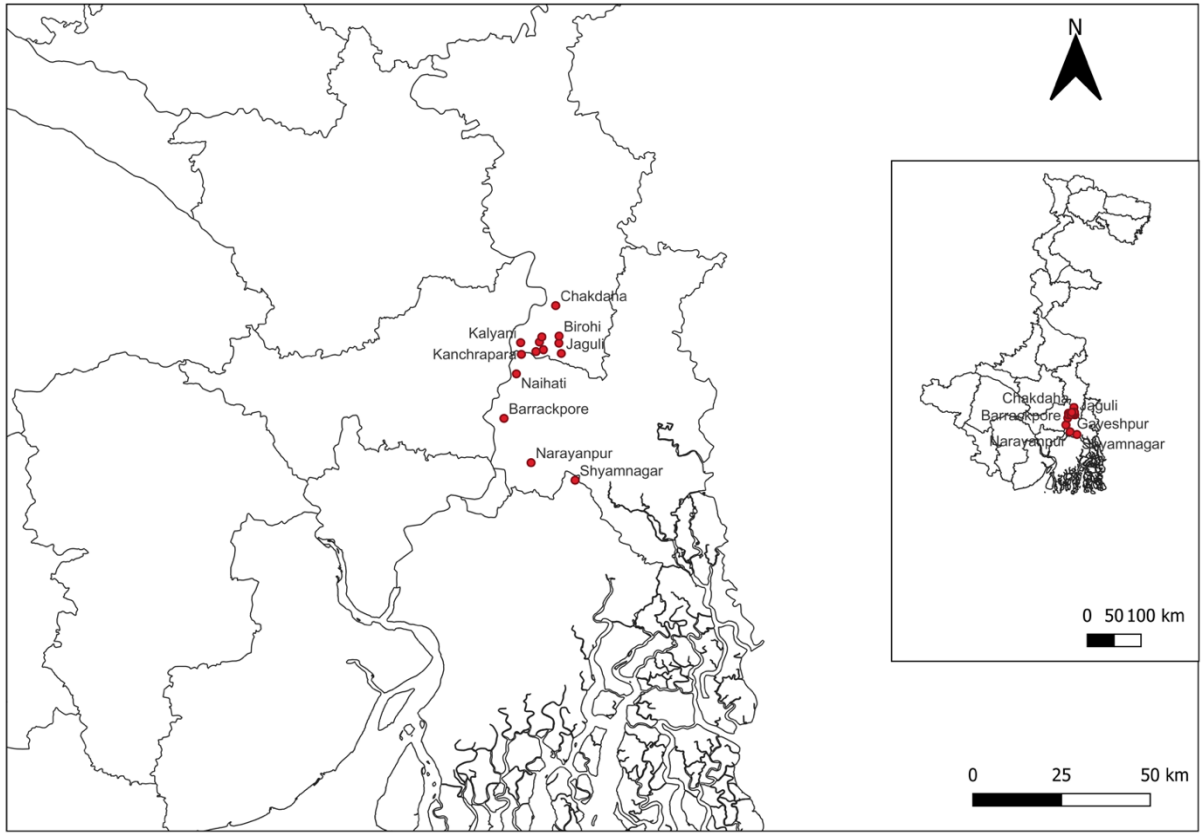

***Figure. 1.*** *The figure presents a map illustrating the 14 study sites across two districts of West Bengal, India. Specifically, in the Nadia district, the study encompassed the following ten locations: Kalyani, Jaguli, Balindi, Chakdaha, Anandanagar, Birohi, Gayeshpur, Kataganj, Narayanpur, and Saguna. In the North 24 Parganas district, the study included four locations: Shyamnagar, Barrackpore, Naihati, and Kanchrapara.*

Free-ranging dogs in these areas subsist primarily by scavenging domestic refuse, and supplement their diets with food obtained from open restaurants, tea shops, meat vendors, and direct feeding by local residents. For a detailed account of resource availability and usage by dogs in these regions, see Bhattacharya and Bhadra (2021).

The study population consisted primarily of adult free-ranging dogs, with only a small number of pups and juveniles being present in the group condition. A total of 650 unique individuals were sampled across two distinct social contexts:

 **Group Condition:** Defined as three or more adult dogs observed within a 5-meter radius. A total of 550 dogs were recorded across 139 distinct group observations.

 **Solitary Condition:** Defined as a single adult dog with no conspecifics within the visual range of the experimenter at the time of observation. This ensured minimal social influence on the focal individual. A total of 145 solitary dogs were sampled.

To prevent resampling, individual dogs were identified based on distinctive coat patterns, scars, and other unique physical features. Sampling sites were rotated, and no site was revisited on the same day to reduce the likelihood of repeated observations.

## 2.2 Experimenters

To minimize observer and experimenter bias, all trials were conducted by pairs of trained experimenters (N = 6; 3 males and 3 females). Each pair consisted of:

**E1 (Primary Experimenter):** Responsible for food presentation and gaze manipulation.

**E2 (Recorder):** Managed video documentation using a hand-held device and ensured unobtrusive observation.

All experimenters received prior training on the standardized protocol. Interactions with dogs were restricted solely to those required by the experiment, maintaining consistent and neutral behaviour across trials.

## 2.3 Experimental Procedure

Each trial was structured into three sequential phases:

### i. Approach Phase (40 seconds):

E2 approached the focal dog(s) at a calm and consistent pace. Upon reaching a 2-meter distance, E2 initiated a 10-second bout of positive vocalizations ("aye-aye-aye") to capture attention (cf. Bhattacharjee et al., 2017), followed by a 30-second settling period. During this time, E1 silently moved into position for the subsequent phase.

### ii. Auditory Cue (2 seconds):

E1 generated a standardised auditory cue by crushing a biscuit packet twice (~2 seconds). This action mimicked a familiar food-related sound to enhance ecological validity.

### iii. Experimental Condition Execution (3 minutes):

One of three randomized conditions was presented:

**GZ_ET (Gazing while Eating):** E1 consumed biscuits while maintaining direct eye contact with the dog(s). In group settings, gaze was alternated among individuals.

**NGZ_ET (No Gazing while Eating):** E1 consumed biscuits while looking away (towards the horizon), avoiding all eye contact.

**GZ_NE (Gazing without Eating):** E1 took a visible bite of a biscuit but withheld further food presentation, maintaining direct eye contact for the duration.

Randomization of conditions controlled for sequence effects across trials.

**2.4 Behavioural Coding and Definitions**

All sessions were recorded and later analysed using a predefined ethogram (adapted from Banerjee & Bhadra, 2021). Behaviours were categorised into active or passive begging based on proximity:

**Active Begging:** Behaviour performed within one body length (BL ≈ 3 ft) of E1.

**Passive Begging:** Behaviour performed within one to three BL of E1, indicating an attentional and food-motivated state without immediate approach.

A behaviour was classified as "begging" only if the dog was alert and not engaged in sleep.

***Table 1.*** *Categorization of begging behaviours exhibited by free-ranging dogs during the experiment. Behaviours were grouped based on functional similarities.*

| Category | Behaviours |
|---|---|
| Posture/Attention | Alert posture, Watching E1 |
| Locomotion | Ambling, walking, trotting, or running towards E1, Approach towards E1 |
| Foraging | Eating fallen biscuit crumbs, Food searching |
| Physical Contact | Jumping on E1, Nudging E1, Pawing (using left or right paw), Mock biting E1, Rubbing head/body on E1, Licking E1 |
| Other | Encircling/walking around E1, Howling, Tail wagging, Salivating, Sniffing E1 |

Begging frequency and duration were extracted from the three-minute experimental phase using frame-by-frame behavioural analysis.

## 2.5 Statistical Analysis

Given the non-normal distribution of behavioural proportion data, all analyses were conducted using non-parametric methods in R v4.2.0 with a significance threshold of α = 0.05. Wilcoxon rank-sum tests were used to compare begging behaviour across (a) social context (solitary vs. group), (b) sex of the dog (male vs. female) and (c) the sex of the experimenter. Kruskal-Wallis rank-sum tests assessed differences in begging behaviour across experimental conditions (GZ_ET, GZ_NE, NGZ_ET). This test was chosen to account for the non-parametric nature of the data.

## 3. Results

### 3.1 Effect of Social Context on Begging Behaviour

Wilcoxon rank-sum tests revealed significant effects of social context on begging behaviour (Figure 2). Dogs observed under solitary conditions exhibited a significantly higher proportion of total begging behaviour than those observed in groups (W = 27,286, $p < 0.001$). This difference was primarily driven by an increased proportion of passive begging among solitary dogs (W = 25,026, $p < 0.001$). In contrast, no significant difference was observed between solitary and group dogs in terms of active begging (W = 38,463, $p > 0.05$).

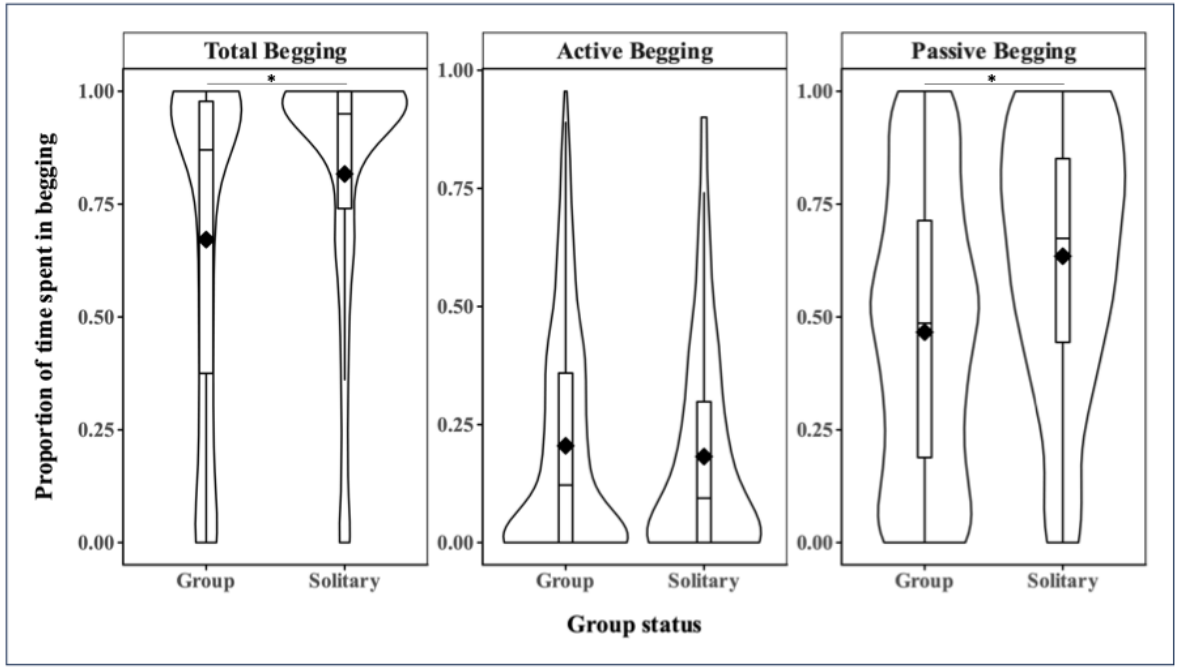

**Figure 2.** *Violin plots showing the proportion of time spent in total, active, and passive begging behaviours by free-ranging dogs, categorized by social context (Group vs. Solitary). Boxplots indicate the median and interquartile range; diamonds represent the mean.*

## 3.2 Sex Differences in Begging Behaviour

Sex-based differences in begging behaviour were also observed (Figure 3). Female dogs showed a significantly higher proportion of total begging than males (W = 58,210, *p* < 0.001), driven largely by elevated levels of passive begging (W = 57,995, *p* < 0.001). No significant sex difference was found in the proportion of active begging (W = 54,930, *p* > 0.05).

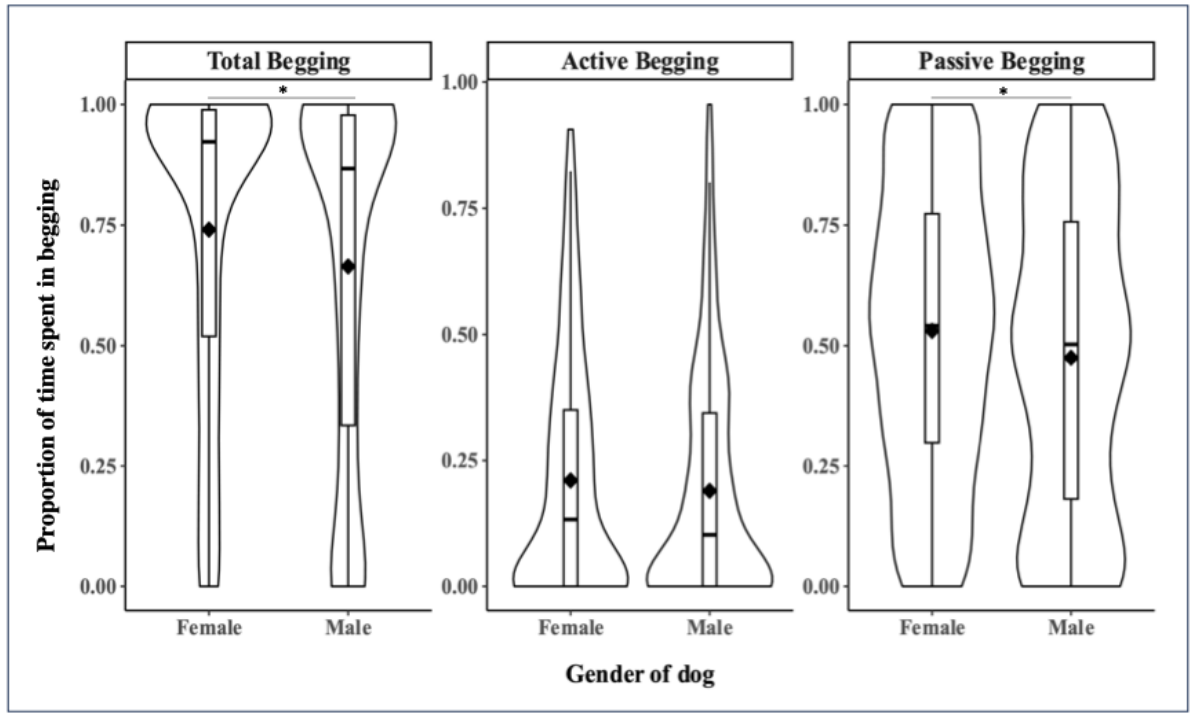

***Figure 3.*** *Violin plots depicting the proportion of time spent in total, active, and passive begging behaviours by male and female free-ranging dogs.*

### 3.3 Effect of Experimental Condition on Begging Behaviour

Kruskal-Wallis rank-sum tests were used to assess differences in begging behaviour across three experimental conditions: Gazing with Eating (GZ_ET), Gazing without Eating (GZ_NE), and No Gazing with Eating (NGZ_ET) (Figure 4). No significant differences were found across conditions for total begging ($\chi^2 = 4.08$, df = 2, $p > 0.05$) or passive begging ($\chi^2 = 2.49$, df = 2, $p > 0.05$). However, a significant difference was observed in active begging ($\chi^2 = 8.63$, df = 2, $p < 0.05$), indicating that active begging behaviour varied depending on the nature of the human cue.

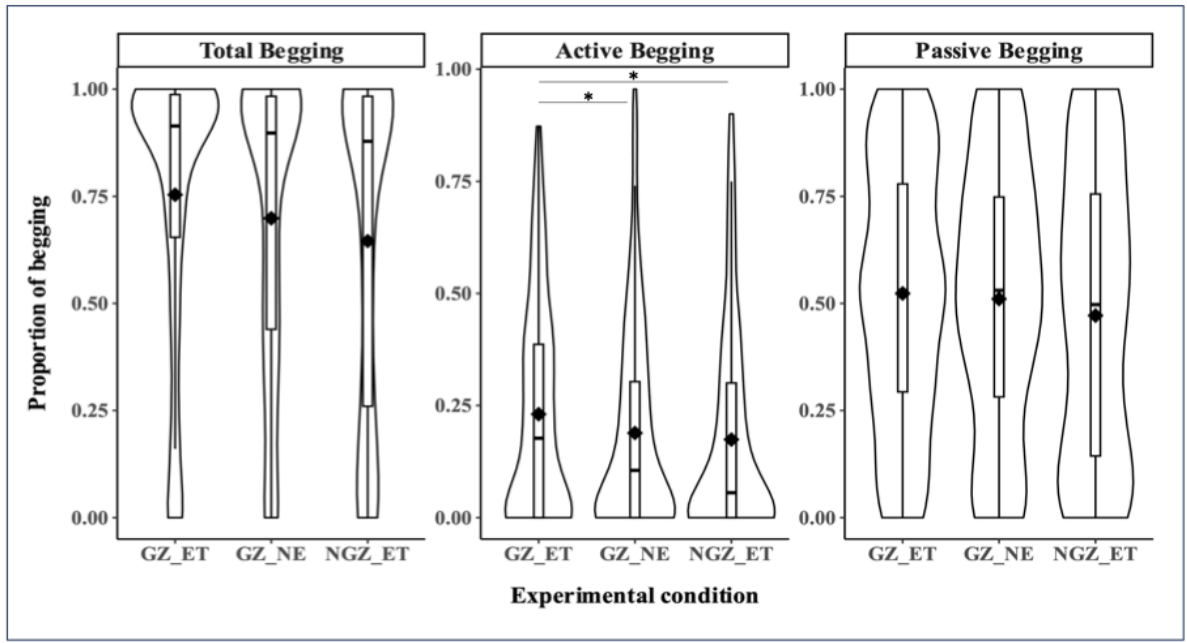

***Figure 4.*** *Violin plots representing the proportion of time spent in total, active, and passive begging behaviours across the three experimental conditions: Gazing with Eating (GZ_ET), Gazing without Eating (GZ_NE), and No Gazing with Eating (NGZ_ET).*

## 3.4 Effect of Experimenter Gender on Begging Behaviour

Begging behaviour also varied with the gender of the experimenter (Figure 5). A significant difference was observed in active begging, with dogs exhibiting higher levels of active begging toward female experimenters (W = 58,088, $p < 0.001$). However, no significant differences were detected in passive begging (W = 52,060, $p > 0.05$) or total begging (W = 56,184, $p > 0.05$) based on experimenter gender.

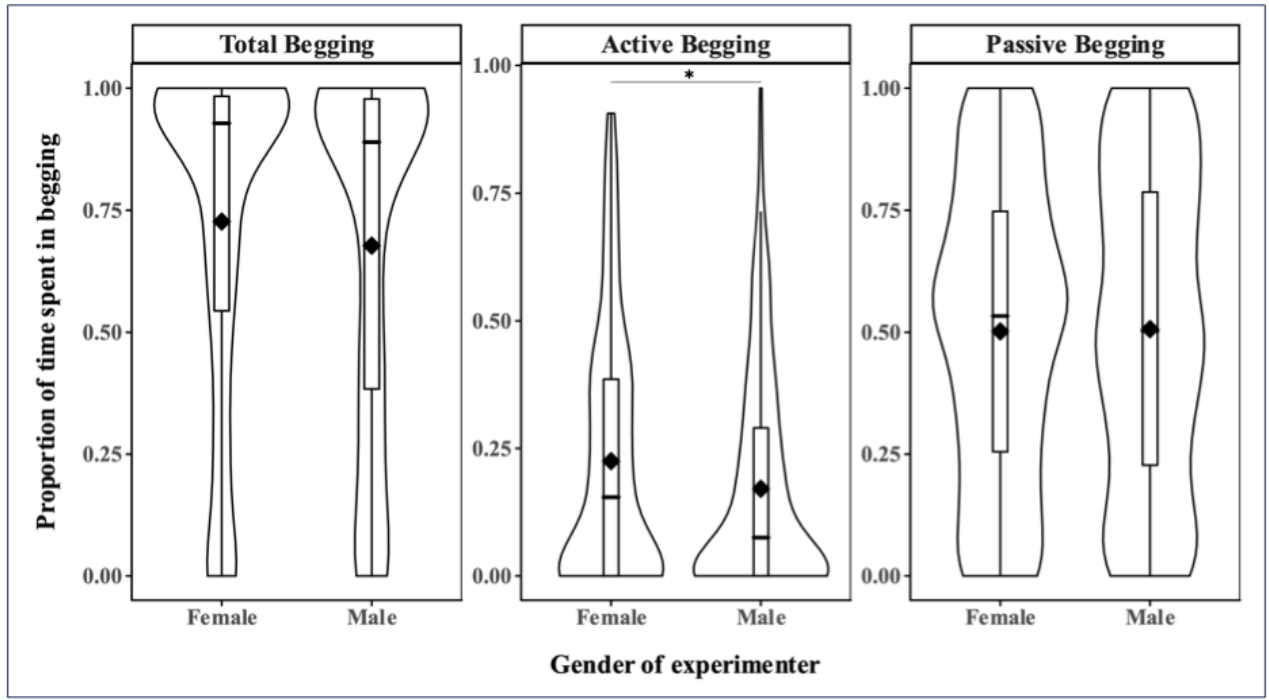

***Figure 5.*** *Violin plots illustrating the proportion of time spent in total, active, and passive begging behaviours when approached by either male or female experimenters.*

## 4. Conclusion

This study examined the nuances of begging strategies in free-ranging dogs, focusing on how social context, the dog's sex, experimental manipulations of human gaze and food availability, and the experimenter's gender influenced total, passive, and active begging. The results reveal a context-sensitive behavioural repertoire, highlighting how dogs flexibly adjust their solicitation strategies based on both social and environmental cues. These findings contribute to a broader understanding of canine social cognition and the mechanisms shaping human–animal interactions in anthropogenic environments.

One key finding was the influence of social context. Dogs in solitary conditions exhibited significantly more total and passive begging than those in groups. This pattern suggests strategic behavioural modulation in response to perceived competition: a solitary dog may perceive a greater probability of obtaining food without interference, thereby favouring lower-effort, passive strategies (Majumder et al., 2014). In contrast, group settings introduce competitive dynamics, potentially reducing the effectiveness—or perceived payoff—of passive displays (Giraldeau & Caraco, 2000; Horn et al., 2012). Interestingly, social context did not significantly affect active begging, suggesting that more effortful, direct behaviours such as pawing or approaching may be governed by individual motivation or perceived opportunity, rather than the immediate presence of conspecifics (Horn et al., 2013).

Sex-based differences also emerged, particularly for passive begging. Female dogs engaged in more total and passive begging than males, consistent with potential hormonal influences or sex-specific behavioural strategies shaped by social experience (Trut, 2000). This may reflect a broader pattern observed across taxa, where females display more socially oriented or affiliative behaviours (Amiot & Bastian, 2017; Cronin, 2012). However, as with social context, sex did not significantly affect active begging, suggesting a behavioural convergence across sexes in high-effort solicitation contexts.

The experimental manipulations of human gaze and food presence revealed more targeted effects. While total and passive begging were unaffected, active begging increased significantly when dogs were subject to human gazing during food consumption. This supports the idea that mutual gaze functions as a salient social cue, enhancing communicative engagement and encouraging solicitation behaviours (Savalli et al., 2016; Bhattacharjee et al., 2017b). Sustained eye contact may act as a permissive signal, increasing the likelihood that dogs will approach or initiate high-effort behaviours. These interactions may be reinforced by physiological mechanisms such as oxytocin-mediated bonding, previously linked to eye contact in dog–human dyads (Bhattacharjee, Sau, et al., 2017).

Remarkably, direct gaze alone—even in the absence of food—elicited begging behaviours, underscoring its communicative potency. Dogs may interpret eye contact as an attentional or affiliative cue, rather than simply a precursor to feeding (Wallis et al., 2015; Byosiere et al., 2023). Conversely, when gaze was absent despite the presence of food, active begging did not increase. This highlights that food availability alone is insufficient to drive high-effort solicitation; rather, dogs rely on attentional cues to assess social opportunity. These findings support prior work indicating that dogs discriminate between gaze directed toward them versus toward inanimate stimuli, adjusting behaviour accordingly (Kaminski & Nitzschner, 2013; Oláh et al., 2017).

We also observed a significant effect of the experimenter's gender on active begging. Dogs exhibited more direct solicitation towards female experimenters, suggesting differential responsiveness based on perceived social or emotional cues. This aligns with findings that dogs can distinguish between male and female humans and often show greater affiliative behaviour toward females, who may engage more in nurturing or communicative interactions (Kubinyi et al., 2017; Powell et al., 2021). Attachment theory provides a useful framework here: prior experiences and reinforcement histories likely shape how dogs interpret and respond to gendered human presence (Svartberg & Forkman, 2002; Morgan & O'Byrne, 2023). Dogs may have learned to associate female humans with greater likelihood of food or attention, particularly in domestic or caregiving contexts (Kubinyi et al., 2009; Turcsán et al., 2020), thereby prompting increased active begging in their presence (Pirrone et al., 2015; Boonhoh et al., 2023).

Taken together, these findings underscore the context-dependent flexibility of canine begging behaviour. Passive begging appears more sensitive to relatively stable social features, such as the dog's sex and social context, while active begging is modulated more strongly by immediate social cues, such as gaze direction and the gender of the human observer. This behavioural divergence suggests a functional distinction between low-effort and high-effort strategies, with the latter more tightly linked to social cognition and real-time assessment of human responsiveness.

In conclusion, this study highlights the behavioural sophistication of free-ranging dogs in navigating human-dominated environments. Their capacity to interpret human gaze, adjust to social conditions, and respond differentially to human sex reveals a high degree of social intelligence and behavioural plasticity. These findings contribute to the growing literature on the co-evolution of interspecies communication and underscore the importance of considering both individual and contextual factors in shaping animal behaviour (Lemma et al., 2022). Future work may explore how individual learning histories, group dynamics, hormonal influences, or neurocognitive profiles shape such strategies, further deepening our understanding of canine social cognition in naturalistic settings.

**References**

Amiot CE & Bastian B. 2017. Solidarity With Animals: Assessing a Relevant Dimension of Social Identification With Animals. *Plos One* 12: e0168184.

Bhattacharjee D, Sau S, Das J & Bhadra A. 2017a. Free-Ranging Dogs Prefer Petting Over Food in Repeated Interactions With Unfamiliar Humans. *Journal of Experimental Biology*.

Bhattacharjee D, Nikhil Dev N, Gupta S, Sau S, Sarkar R, Biswas A, Banerjee A, Babu D, Mehta D & Bhadra A. 2017b. Free-ranging dogs show age related plasticity in their ability to follow human pointing. *PLoS ONE* 12.

Bhattacharjee D, Dasgupta S, Biswas A, Deheria J, Gupta S, Nikhil Dev N, Udell M & Bhadra A. 2017c. Practice makes perfect: familiarity of task determines success in solvable tasks for free-ranging dogs (Canis lupus familiaris). *Animal Cognition* 20: 771–776.

Bhattacharjee D & Bhadra A. 2021. Response to short-lived human overcrowding by free-ranging dogs. *Behavioral Ecology and Sociobiology* 75.

Boonhoh W, Wongtawan T, Sriphavatsarakom P, Waran N & Boonkaewwan C. 2023. Factors Associated With Pet Dog Behavior in Thailand. *Veterinary World*: 957–964.

Bräuer J, Call J & Tomasello M. 2005. All great ape species follow gaze to distant locations and around barriers. *Journal of Comparative Psychology* 119: 145.

Bugnyar T, Stöwe M & Heinrich B. 2004. Ravens, Corvus corax, follow gaze direction of humans around obstacles. *Proceedings of the Royal Society of London. Series B: Biological Sciences* 271: 1331–1336.

Byosiere S, Mundry R, Range F & Virányi Z. 2023. Selective Responding to Human Ostensive Communication Is an Early Developing Capacity of Domestic Dogs. *Developmental Science* 26.

Cronin KA. 2012. Prosocial Behaviour in Animals: The Influence of Social Relationships, Communication and Rewards. *Animal Behaviour* 84: 1085–1093.

Daniels TJ & Bekoff M. 1989. *Population and Social Biology of Free-Ranging Dogs, Canis familiaris.*

Emery NJ. 2000. The eyes have it: the neuroethology, function and evolution of social gaze. *Neuroscience & biobehavioral reviews* 24: 581–604.

Giraldeau LA & Caraco T. 2000. *Social foraging theory.* Princeton University Press.

Hare B & Tomasello M. 2005. Human-like social skills in dogs? *Trends in Cognitive Sciences* 9: 439–444.

Horn L, Huber L & Range F. 2013. The Importance of the Secure Base Effect for Domestic Dogs – Evidence From a Manipulative Problem-Solving Task. *Plos One* 8: e65296.

Horn L, Range F & Huber L. 2012. Dogs' Attention Towards Humans Depends on Their Relationship, Not Only on Social Familiarity. *Animal Cognition* 16: 435–443.

Kaminski J, Call J & Tomasello M. 2004. Body orientation and face orientation: Two factors controlling apes' begging behavior from humans. *Animal Cognition* 7: 216–223.

Kaminski J & Nitzschner M. 2013. Do dogs get the point? A review of dog-human communication ability. *Learning and Motivation* 44: 294–302.

Kubinyi E, Bence M, Koller D, Wan M, Pergel E, Rónai Z, Sasvári-Székely M & Miklósi Á. 2017. Oxytocin and Opioid Receptor Gene Polymorphisms Associated With Greeting Behavior in Dogs. *Frontiers in Psychology* 8.

Kubinyi E, Turcsán B & Miklósi Á. 2009. Dog and Owner Demographic Characteristics and Dog Personality Trait Associations. *Behavioural Processes* 81: 392–401.

Lemma M, Doyle RE, Alemayehu G, Mekonnen MM, Kumbe A & Wieland B. 2022. Using Community Conversations to Explore Animal Welfare Perceptions and Practices of Rural Households in Ethiopia. *Frontiers in Veterinary Science* 9.

Majumder S Sen, Bhadra A, Ghosh A, Mitra S, Bhattacharjee D, Chatterjee J, Nandi AK & Bhadra A. 2014. To be or not to be social: Foraging associations of free-ranging dogs in an urban ecosystem. *Acta Ethologica* 17: 1–8.

Miklósi Á, Pongrácz P, Lakatos G, Topál J & Csányi V. 2005. A comparative study of the use of visual communicative signals in interactions between dogs (Canis familiaris) and humans and cats (Felis catus) and humans. *Journal of comparative psychology* 119: 179.

Miklósi Á. 2014. *Dog behaviour, evolution, and cognition*. oUp Oxford.

Morgan S & O'Byrne DA. 2023. How Autism Assistance Canines Enhance the Lives of Autistic Children. *Inquiry the Journal of Health Care Organization Provision and Financing* 60.

Nagasawa M, Mitsui S, En S, Ohtani N, Ohta M, Sakuma Y, Onaka T, Mogi K & Kikusui T. 2015. Oxytocin-gaze positive loop and the coevolution of human-dog bonds. *Science* 348: 333–336.

Oláh K, Topál J, Kovács K, Kis A, Koller D, Park SY & Virányi Z. 2017. Gaze-Following and Reaction to an Aversive Social Interaction Have Corresponding Associations With Variation in the OXTR Gene in Dogs but Not in Human Infants. *Frontiers in Psychology* 8.

Pirrone F, Pierantoni L, Mazzola S, Vigo D & Albertini M. 2015. Owner and Animal Factors Predict the Incidence Of, and Owner Reaction Toward, Problematic Behaviors in Companion Dogs. *Journal of Veterinary Behavior* 10: 295–301.

Poucke E Van, Höglin A, Jensen P & Roth LS V. 2021. Breed Group Differences in the Unsolvable Problem Task: Herding Dogs Prefer Their Owner, While Solitary Hunting Dogs Seek Stranger Proximity. *Animal Cognition* 25: 597–603.

Povinelli DJ, Eddy TJ, Hobson RP & Tomasello M. 1996. What young chimpanzees know about seeing. *Monographs of the society for research in child development*: i–189.

Powell L, Stefanovski D, Siracusa C & Serpell JA. 2021. Owner Personality, Owner-Dog Attachment, and Canine Demographics Influence Treatment Outcomes in Canine Behavioral Medicine Cases. *Frontiers in Veterinary Science* 7.

Savalli C, Resende B & Gaunet F. 2016. Eye contact is crucial for referential communication in pet dogs. *PLoS ONE* 11.

Somppi S, Törnqvist H, Topál J, Koskela A, Hänninen L, Krause CM & Vainio O. 2017. Nasal oxytocin treatment biases dogs' visual attention and emotional response toward positive human facial expressions. *Frontiers in Psychology* 8: 1854.

Svartberg K & Forkman B. 2002. Personality Traits in the Domestic Dog (Canis Familiaris). *Applied Animal Behaviour Science* 79: 133–155.

Téglás E, Gergely A, Kupán K, Miklósi Á & Topál J. 2012. Dogs' Gaze Following Is Tuned to Human Communicative Signals. *Current Biology* 22: 209–212.

Topál J, Kis A & Oláh K. 2014. Dogs' Sensitivity to Human Ostensive Cues: A Unique Adaptation? A Unique Adaptation? In: *The Social Dog: Behavior and Cognition*. Elsevier Inc., 319–346.

Trut LN. 2000. Early canid domestication: the farm-fox experiment. *Scientifur* 24: 124.

Turcsán B, Tátrai K, Petró E, Topál J, Balogh L, Egyed B & Kubinyi E. 2020. Comparison of Behavior and Genetic Structure in Populations of Family and Kenneled Beagles. *Frontiers in Veterinary Science* 7.

Udell MAR, Dorey NR & Wynne CDL. 2010. What did domestication do to dogs? A new account of dogs' sensitivity to human actions. *Biological reviews* 85: 327–345.

Wallis L, Range F, Müller CA, Serisier S, Huber L & Virányi Z. 2015. Training for Eye Contact Modulates Gaze Following in Dogs. *Animal Behaviour* 106: 27–35.

Werhahn G, Virányi Z, Barrera G, Sommese A & Range F. 2016. Wolves (Canis lupus) and dogs (Canis familiaris) differ in following human gaze into distant space but respond similar to their packmates' gaze. *Journal of Comparative Psychology* 130: 288.

Zajonc RB. 1965. Social Facilitation: A solution is suggested for an old unresolved social psychological problem. *Science* 149: 269–274.

**Acknowledgements**

The authors would like to acknowledge the Indian Institute of Science Education and Research Kolkata for providing infrastructural support.

**Funding**

SB would like to thank the University Grants Commission, India for providing him doctoral fellowship. The project was partially funded by the Janaki Ammal Award grant BT/HRD/NBA-NWB/39/2020-21 (YC-1), to AB by the Department of Biotechnology, India.

**Author contributions**

SB, SN, TP, AL, AR, HG carried out the field work. SN, TP and KG decoded the videos. SB and SN carried out the data analysis. SB wrote the manuscript. SB conceived the idea. AB helped in finalizing the experimental protocol, supervised the work, acquired funding, reviewed and edited the manuscript.

**Data availability statement**

Data supporting the results will be archived.

**Conflict of Interest Information**

Authors have no conflict of interest.